\documentclass{elsart}
\usepackage{epsfig}
\usepackage{amsmath}
\usepackage{color}
\usepackage{amssymb}
\usepackage{graphicx}
\usepackage{subfigure}
\usepackage{rotating}
\usepackage{setspace,url,hyperref}

\newcommand \be{\begin{equation}}
\newcommand \ba{\begin{eqnarray}}
\newcommand \ee{\end{equation}}
\newcommand \ea{\end{eqnarray}}

\begin{document}

\begin{frontmatter}


\title{Exploring self-similarity of complex cellular networks:
the edge-covering method with simulated annealing and log-periodic sampling}
\author[ecust1,ecust2,nice]{\small{Wei-Xing Zhou}},
\author[ecust1]{\small{Zhi-Qiang Jiang}},
\author[ETH,nice]{\small{Didier Sornette}\thanksref{EM}}
\address[ecust1]{School of Business and Research Center of Systems Engineering, East
China University of Science and Technology, Shanghai 200237, China}
\address[ecust2]{School of Science, East
China University of Science and Technology, Shanghai 200237, China}
\address[ETH]{D-MTEC, ETH Zurich, CH-8032 Zurich, Switzerland}
\address[nice]{LPMC, CNRS UMR 6622 and Universit\'e de Nice-Sophia
Antipolis, 06108 Nice
Cedex 2, France}
\thanks[EM]{Corresponding author. {\it E-mail address:}\/
dsornette@ethz.ch}

\begin{abstract}
Song, Havlin and Makse (2005) have recently used a version of the box-counting
method, called the node-covering method, to quantify the self-similar
properties of 43 cellular networks: the minimal number $N_V$ of boxes of size $\ell$
needed to cover all the nodes of a cellular network
was found to scale as the power law $N_V \sim (\ell+1)^{-D_V}$ with
a fractal dimension $D_V=3.53\pm0.26$. We propose a new box-counting
method based on edge-covering, which outperforms the node-covering
approach when applied to strictly self-similar model networks, such as the
Sierpinski network. The minimal number $N_E$ of boxes of size $\ell$
in the edge-covering method is obtained with the simulated annealing
algorithm. We take into account the possible discrete scale symmetry
of networks (artifactual and/or real), which is visualized
in terms of log-periodic oscillations in the
dependence of the logarithm of $N_E$  as a function of the logarithm
of $\ell$. In this way, we are able to remove the bias of the estimator of the fractal
dimension, existing for finite networks. With this new methodology, we
find that $N_E$ scales with respect to $\ell$ as a
power law $N_E \sim \ell^{-D_E}$ with $D_E=2.67\pm0.15$ for the 43
cellular networks previously analyzed by Song, Havlin and Makse (2005).
Bootstrap tests suggest that the analyzed cellular networks may
have a significant log-periodicity qualifying
a discrete hierarchy with a scaling ratio close to $2$.
In sum, we propose that our method
of edge-covering with simulated annealing and log-periodic sampling
minimizes the significant bias in the determination of
fractal dimensions in log-log regressions.

\end{abstract}


\end{frontmatter}

\section{Introduction}

In recent years, the study of complex networks have attracted
extensive interest, covering biological, social, information, and
technological systems
\cite{Albert-Barabasi-2002-RMP,Newman-2003-SIAMR,Dorogovtsev-Mendes-2003}.
Most complex networks exhibit small-world properties
\cite{Watts-Strogatz-1998-Nature} and are scale free in the sense
that the distribution of degrees has power-law tails
\cite{Barabasi-Albert-1999-Science}. Note that, as stressed recently
by Keller \cite{Keller-2005-BE}, the existence of power law
distribution in many complex networks is a re-discovery in this
field of previously well-known mechanisms (see for instance chapters
14 and 15 of \cite{Sornette-2004} which describes many mechanisms
for power laws which have been described in several often quite
different scientific contexts). In addition, many real networks have
modular structures or communities \cite{Newman-2004-EPJB} expressing
their underlying functional modules. The fourth intriguing feature
of some (not all) real networks reported recently is the
self-similarity of the topology
\cite{Song-Havlin-Makse-2005-Nature}, characterized by a fractal
dimension \cite{Mandelbrot-1983}, and the scale invariance of degree
distribution after coarse graining processes \cite{Kim-2004-PRL}.

The fractal nature of a self-similar network can be revealed by
utilizing the well-known box-counting method
\cite{Mandelbrot-1983,Feder-1988}. Song, Havlin, and Makse
\cite{Song-Havlin-Makse-2005-Nature} have adopted a node-covering
method, in which boxes of size $\ell_B=\ell+1$ are used to cover all
the nodes of a network, where $\ell$ is the diameter of the
subnetwork enclosed in the box, and the minimum number of
node-covering boxes, $N_V(\ell_B)$, is determined for each $\ell_B$.
If the network is self-similar, $N_V(\ell_B)$ scales with respect to
$\ell_B$ as a power law
\begin{equation}\label{Eq:Nv:ell}
     N_V \sim 1/(\ell+1)^{D_V}~,
\end{equation}
where $D_V$ is the fractal dimension of the network.

It is natural to raise the question of what is the underlying
mechanism for a network to evolve into a self-similar structure
\cite{Strogatz-2005-Nature}. An empirical study reports that two
genetic regulatory networks, which are self-similar scale-free
networks, are not assortative (i.e., cannot be separated into groups
according to kind), a result confirmed by numerical simulations
\cite{Yook-Radicchi-Meyer-Ortmanns-2005-PRE}. However, this claim
was falsified by Song, {\it{et al.}}, who showed that the
self-similar structure in complex networks was due to the repulsion
between hubs \cite{Song-Havlin-Makse-2006-NP}. In addition,
self-similar networks have been shown to have the same fractal scaling
as thei skeleton \cite{Goh-Salvi-Kahng-Kim-2006-PRL}. An alternative
node-covering method was proposed based on the skeleton of the
network under investigation \cite{Kim-Goh-Salvi-Oh-Kahng-Kim-2006-XXX}.

Another important symmetry associated with scale invariance and
fractals is discrete scale invariance (DSI) \cite{Sornette-1998-PR},
which expresses the property that the fractal is self-similar only
with respect to magnification factors which are integer powers of a
preferred scaling ratio $\lambda$: when a fractal possesses the
property of DSI, it is self-similar only under magnification with
factors $\lambda$, $\lambda^2$, $\lambda^3$, $\lambda^4$, and so on.
The observable hallmark of discrete scale invariance is
log-periodicity in the scaling of the observables as a function of
scale or control parameter(s). For instance, significant
log-periodic oscillations are observed in the degree distributions
\cite{Suchecki-Holyst-2005-APPB} and energy distributions
\cite{Grana-Pinasco-2006-XXX} of model networks. However, there is
no direct evidence, yet, showing the presence of log-periodicity in
the topological structure of self-similar networks observed in
Nature or in social structures.

In this paper, we first study a model network, known as the Sierpinski gasket
in the context of fractals, which is exactly self-similar with
build-in log-periodicity. We propose a novel box-counting method
based on edge-covering tiling, which significantly outperforms the
node-covering method in the determination of fractal dimensions.
A new method for detecting discrete scale invariance is also proposed
and tested on the Sierpinski network. We then combine these two methods
to obtain an unbiased estimator of the fractal dimension of self-similar
networks. We apply this methodology
to the forty three cellular networks investigated by Song, {\it{et
al.}} \cite{Song-Havlin-Makse-2005-Nature} and find
significant differences. In particular, it seems that previous
results have been significantly biased upwards. We also suggest
the existence of a weak discrete dichotomous hierarchical structure
in many of the analyzed networks.

\section{A family of exactly self-similar networks}
\label{s1:Sierpinski}

\subsection{The construction of Sierpinski networks}
\label{s2:Sierpinski}

\begin{figure}[htb]
\begin{center}
\includegraphics[width=7cm]{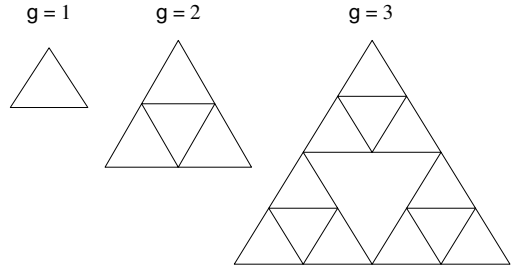}
\end{center}
\caption{Construction of Sierpinski networks. A
Sierpinski network of generation $g$ is denoted $\Delta_g$.
The three networks shown in this figure are thus $\Delta_1$, $\Delta_2$ and
$\Delta_3$.} \label{Fig:Sierpinski}
\end{figure}

Let us first recall how to construct the family of Sierpinski
triangle networks. The initiator is an equilateral triangle of unit
length. Two replicas of the initiator are placed near the initiator
to form an equilateral triangle, of the same form as the generator
but with side length twice larger. Starting from the initiator of
generation $g=1$, we thus obtain in the first iteration of the
construction the generation $g=2$ formed of three initiators. The
first three generations of the iterative construction of Sierpinski
networks is depicted in Fig.~\ref{Fig:Sierpinski}.

It is well-known that Sierpinski networks are self-similar. For a
Sierpinski network of generation $g$, the number of nodes is
\begin{equation}
  n_g = \frac{3^g+3}{2}~,
  \label{Eq:ng}
\end{equation}
which is obtained of the recursion relation
$n_{g+1} = 3 n_g -3$ with $n_{g=1}=3$. The
number of edges is simply
\begin{equation}
  m_g = 3^g~.
  \label{Eq:mg}
\end{equation}
For instance, a Sierpinski network of generation $g=6$ has $n_g =
366$ nodes and
$m_g = 729$ edges.

\subsection{Box-counting method by covering nodes}
\label{s2:NCBC}

\begin{figure}[htb]
\begin{center}
\includegraphics[width=8cm]{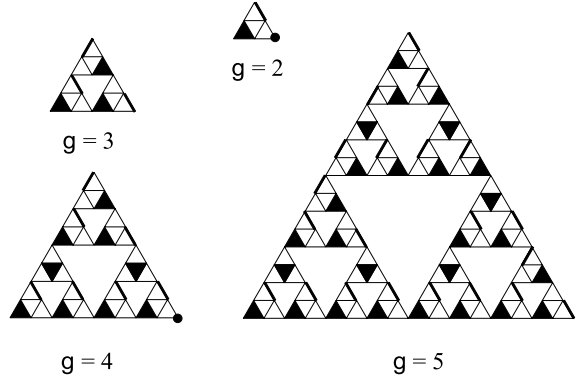}
\caption{Iterative algorithm for tiling Sierpinski networks
$\Delta_g$ for $g=2$, $3$, $4$ and $5$ with boxes of size $\ell=1$.
The ``boxes'' are the triangle initiator $\Delta_1$, an edge
of unit length with its two nodes (represented by as
thick segments in the figure) and an individual node (represented
as a small filled circle in the figure).
}
\label{Fig:Tiling}
\end{center}
\end{figure}

We first adapt the node-covering method used by Song, Havlin, and
Makse \cite{Song-Havlin-Makse-2005-Nature} to cover the nodes of a
Sierpinski network $\Delta_g$. Let us start with a box of size
$\ell_B=\ell=1$. Actually, three geometrical patterns should be
considered which are of box size $\ell=1$: (i) the triangle initiator
$\Delta_1$, (ii) an edge of unit length with two nodes, and (iii) an individual
node. We use these three geometrical patterns to cover the nodes of
Sierpinski networks. Figure \ref{Fig:Tiling} shows the obtained
coverings of $\Delta_g$ for $g=2$, $3$, $4$ and $5$ with boxes of
size $\ell=1$. For instance, the optimal node-covering configuration
of $\Delta_2$ is obtained with just one triangle initiator
$\Delta_1$, one edge and one individual node. The optimal
node-covering configuration of $\Delta_3$ uses three triangle
initiators $\Delta_1$, three edges and zero individual nodes. And so
on.

For a given network, its node-covering structure includes
$N_3$ triangle initiators $\Delta_1$, $N_2$
edges, and $N_1$ nodes, giving a total number of covering boxes equal to
$N_V=N_1+N_2+N_3$. Note that the number of nodes is
given by $n_g=3N_3+2N_2+N_1$ because the optimal
node-covering patterns involve no overlaps. For $g=1$, we have obviously
$N_V(1)=1$. For $g\ge2$, each Sierpinski network $\Delta_g$ can
be viewed as made of $\Delta_2$ elements. For each $\Delta_2$
in a given $\Delta_g$, there is only one covering box which is a
$\Delta_1$. For two adjacent $\Delta_2$'s, there are no additional
$\Delta_1$ boxes.
Therefore, the maximal number of boxes $\Delta_1$ used in the covering is
\begin{equation}
  N_3'=3^{g-2}~.
  \label{Eq:N3}
\end{equation}
These $N_3'$ $\Delta_1$ boxes cover $3N_3'=3^{g-1}$ nodes. Therefore,
the minimal number of nodes
which are not covered by $\Delta_1$ boxes is $n_g-3N_3' =
\frac{3^g+3}{2}-3^{g-1} = \frac{3^{g-1}+3}{2}$. For this minimum number
of still uncovered nodes, the maximum number
of covering boxes that are edges (which we can refer to as ``Min-Max'') is
\begin{equation}
  N_2' = \left[ \frac{n_g-N_3'}{2} \right] = \left[ \frac{3^{g-1}+3}{4}
  \right]= \frac{3^{g-1}-(-1)^{g-1}}{4} + \frac{1+(-1)^{g-1}}{2}~,
  \label{Eq:N2}
\end{equation}
where $[x]$ is the integer part of $x$. Then, the remaining number of isolated
nodes to cover with individual node boxes is
\begin{equation}
  N_1' = n_g-3N_3'-2N_2' = [1-(-1)^{g-1}]/2~,
  \label{Eq:N1}
\end{equation}
It follows immediately that
\begin{equation}
  N_V \ge N_1'+N_2'+N_3' = \frac{7\times 3^{g-2}+4+(-1)^{g}}{4}~.
  \label{Eq:NVge}
\end{equation}
In other words, because the total number of covering boxes is
minimized by using the maximum number of $\Delta_1$'s (since each
cover three nodes), the above construction shows that
$N_1'+N_2'+N_3'$ is the infimum of the minimal number $N_V$ of
covering boxes.

For $g=2,3,4,5$, Fig.~\ref{Fig:Tiling} provides one possible tiling
for each, where the infimum $N_1'+N_2'+N_3'$ is reached, i.e., the
number of boxes is exactly $[{7\times 3^{g-2}+4+(-1)^{g}}]/{4}$.
Together with Eq.~(\ref{Eq:NVge}), we synthesize the above results
under the single expression
\begin{equation}
  N_V = H(g-1.5) \times \frac{7\times 3^{g-2}+4+(-1)^{g}}{4} +
  H(1.5-g)~,
  \label{Eq:NV}
\end{equation}
where $H(x)$ is the Heaviside function. Our numerical calculations show that
this equation holds for all $g>5$ cases that we have investigated.
For the covering
of a Sierpinski network $\Delta_g$ with boxes of size $\ell>1$, we do
not have analytic
expressions for $N_V(\ell)$. However, our numerical simulations show
that $N_V(2^k)=N_E(2^k)$ for $1<k<g$, where the definition of $N_E$ is
given in the next sub-section.

\subsection{Box-counting method by covering edges}
\label{s2:ECBC}

\begin{figure}[htb]
\begin{center}
\includegraphics[width=13cm]{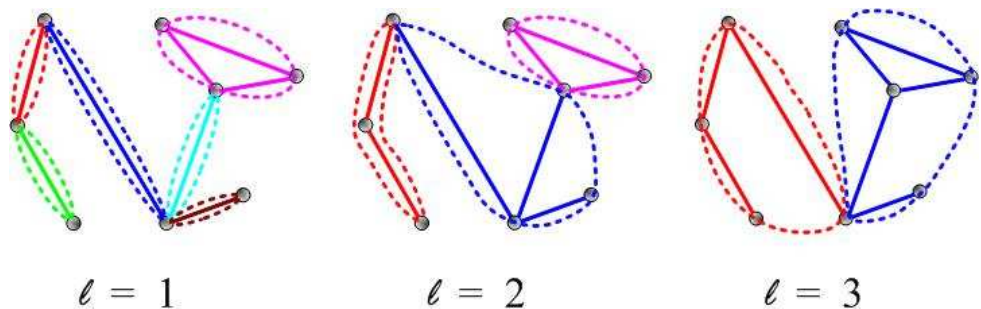}
\caption{(Color online) Illustration of the edge-covering method for
three different ``box'' sizes $\ell=1$, $2$, and $3$. The nodes of
the network are the circles and the edges of the network are shown
as solid lines. The closed dashed lines delineate the sub-sets or
boxes of diameter no larger than $\ell$. For $\ell=1$, two subset
topologies are possible: edges joining two adjacent nodes and a
triplet of nodes forming a triangle which also has diameter $1$. For
larger $\ell$, arbitrary complicated topologies for the
sub-set/boxes are possible as long as their diameter remains smaller
than or equal to $\ell$, as shown in the case $\ell=2$ and $\ell=3$.
The edge-covering shown in the figure was obtained by our simulated
annealing algorithm which provides in these simple cases optimal
solutions.} \label{Fig:EdgeCovering}
\end{center}
\end{figure}

Instead of covering nodes as described in the previous sub-section,
we propose to use boxes in order to cover all the edges.
Mathematically, this problem can be described as follows. Consider
the set $E$ of all the edges of a given network. Let $\{E_i:
i=1,2,\cdots, N_E(\ell)\}$ be a partition of $E$, that is, $E_i
\subseteq E$, $E_i\cap E_j=\Phi$ (null set) for $i\ne j$, and $E =
\cup_{i=1}^{N_E(\ell)} E_i$. The size (or diameter) of $E_i$,
denoted by $d(E_i)$, is defined as the maximum distance among all
possible pairs of nodes of $E_i$ counted as the number of edges to
go from one node to the other of the pair. For a given $\ell$, we
construct an edge-covering by partitioning the set $E$ into subsets
$\{E_i\}$ such that all their diameters $d(E_i)$ are smaller than or
equal to $\ell$ for all $i$'s. Then, the partition $\{E_i:
i=1,2,\cdots, N_E(\ell)\}$  is said to be a covering of the network
with boxes of size $\ell$. We look for the best edge-covering
structures by minimizing $N_E(\ell)$, for each $\ell$. An
illustration of the edge-covering method is shown in
Fig.~\ref{Fig:EdgeCovering} for three different values of $\ell$.

Note that our algorithm for covering edges also automatically covers the nodes
(sometimes with some nodes multiply covered). Thus, the minimal number
$N_E(\ell)$ of boxes of a given size $\ell$ in the edge-covering
method is not smaller than $N_V(\ell)$. Consider a $g$-generation
Sierpinski network and scales $\ell = 2^k$ with
$k=0,1,2,\cdots,g-1$. It is well-known that
\begin{equation}
  N_E(2^k) = 3^{g-1-k}~.
  \label{Eq:NE}
\end{equation}
Therefore, we have
\begin{equation}
  N_E(\ell) = 3^{g-1} \cdot \ell^{-D_E}~,
  \label{Eq:DE}
\end{equation}
where $D_E=\ln(3)/\ln(2)\approx 1.5850$ is the fractal dimension. On
the other hand, the scale invariance symmetry of the Sierpinski
network can be expressed as
\begin{equation}
  N_E(\ell) = 3N_E(2\ell)~.
  \label{Eq:NESI}
\end{equation}
The general solution of this functional equation (\ref{Eq:NESI}) takes
the log-periodic structure
\begin{equation}
  N_E(\ell) \sim \ell^{-D_E}
  \phi\left(\log_2(\ell)\right)~,
  \label{Eq:NEDSI}
\end{equation}
where $\phi(x+1)=\phi(x)$ is a periodic function with unit period.
Here, the preferred scaling ratio of the construction of the
Sierpinski gasket is $\lambda = e^{\ln(2)} =2$, corresponding to a
log-frequency $f=1$ (more generally when the scaling ratio is not
$2$, we have $\phi(x+1/f)=\phi(x)$ and $\lambda = e^{\ln(2)/f}$).

\subsection{Simulated annealing for the node-covering and
edge-covering methods}

According to the definition of fractal dimensions,
we need to determine the minimal number of boxes necessary to cover the nodes or the
edges of a network, as a function of the box scale $\ell$.
In the supplementary materials of
ref.~\cite{Song-Havlin-Makse-2005-Nature}, it was reported that ``the
minimization (of the number of boxes used to cover the network) is
not relevant and any covering gives the same exponent''.
While this may be true in some cases, working with arbitrary
covering structures adds noise to the scaling laws. In order
to get cleaner scaling, we have implemented
the simulated annealing algorithm
\cite{Kirkpatrick-Gelatt-Vecchi-1983-Science,Basu-Frazer-1990-Science}
to obtain covering partitions with a minimum number of covering boxes.

The simulated annealing algorithm is implemented as follows.
For the edge-covering problem, the network is viewed as a set
of edges, which can be partitioned into ``boxes'' or subsets of connected
edges. Starting from a given partition with $C$ boxes of sizes no
larger than $\ell$, we consider three possible moves to transform
the partition into a new one:
\begin{enumerate}
   \item One edge is moved from one box with at least two edges in it to
   another box if the diameters of both new boxes do not exceed $\ell$;
  \item One edge is moved out of one box with at least two edges to form
  a new box consisting of one edge;  and
   \item Two boxes merge to form a new box, a move which is allowed if
the diameter
   of the resulting box is no larger than $\ell$.
\end{enumerate}
At each temperature $T$, we perform $k_1$ edge operations of the
first-type of edge exchanges between pairs of boxes, $k_2$ operations of the
second-type to form new boxes, and $k_3$ operations of
the third-type merging pairs of boxes. An operation is
accepted with probability
\begin{equation}\label{Eq:p:SA}
     p = \left\{
     \begin{array}{lll}
       1                                   &~~&  {\rm{if}}~~C_a\leqslant C_b \\
       \exp\left(-\frac{C_a-C_b}{T}\right) &~~& {\rm{if}}~~C_a>C_b
     \end{array}
     \right.
\end{equation}
where $C_b$ and $C_a$ are the numbers of boxes {\it{before}} and
{\it{after}} an operation. After having performed $k_1+k_2+k_3$
possible operations, the system is cooled down to a lower
temperature $T'=cT$, where $c$ is a constant less than and close to
$1$ (typically $c=0.995\sim0.999$).

For the node-covering method, the simulated annealing procedure is
deduced from the one described above by applying the three types of
operations to nodes rather than to edges. Typical values for $k_1,
k_2$ and $k_3$ are 20000, 5, and 15. The much larger value of $k_1$
is justified by the fact that the
single edge transfers should be much more frequent
than merging two boxes
and splitting one boxes into two. If $k_2$ and $k_3$ were much larger
relative to $k_1$, it would be more difficult
for the optimization to converge to a minimal box number. This
is confirmed by our simulations.

\subsection{Numerical comparison of the nodes-covering and
edges-covering box-counting methods}
\label{s2:Cmp:NV:NE}

\begin{figure}[htb]
\begin{center}
\includegraphics[width=10cm]{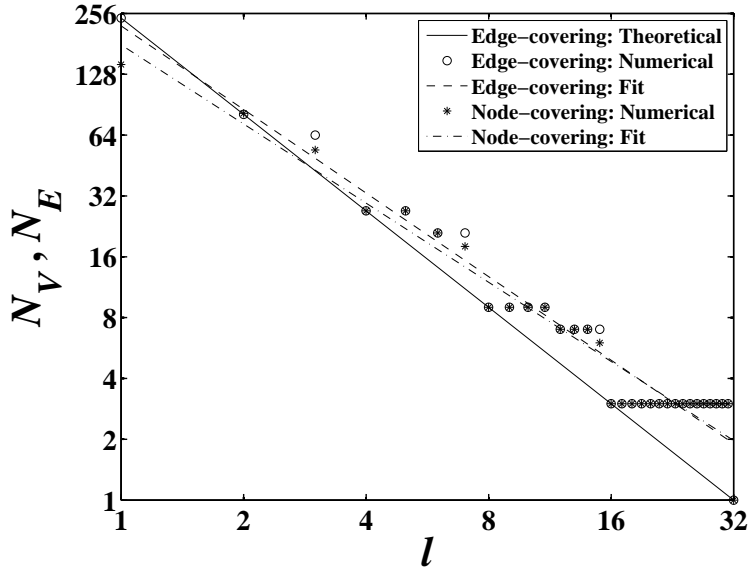}
\end{center}
\caption{Comparison of simulations and theoretical analysis for the
node-covering and edge-covering methods described in previous
sections. The slope of the dotted-dashed (respectively dashed) line
gives the estimation $D_V = 1.30$ (respectively $D_E=1.37$) from the
node-covering (respectively edge-covering) method. The exact
analytical value is $D_f=1.5850$. } \label{Fig:Sierpinski:NVNE}
\end{figure}

We first compare the node-covering method developed in
Ref.~\cite{Song-Havlin-Makse-2005-Nature} and our edge-covering
method by applying them on a Sierpinski network $\Delta_6$ of sixth
generation. We have used the simulated annealing algorithm described
above to search for the minimal numbers $N_V(\ell)$ and $N_E(\ell)$
of covering boxes of size $\ell$ for the two methods. The results
are shown in Fig.~\ref{Fig:Sierpinski:NVNE}. One can observe that
$N_V(1)=143$ predicted by (\ref{Eq:NV}) and $N_E(2^k)$ predicted by
Eq.~(\ref{Eq:NE}) are recovered by the simulations. Our simulations
also confirm that $N_V(\ell)\leqslant N_E(\ell)$, as expected from
the nature of the two methods.

Although the values of $N_V$ and $N_E$ are quite close, their
difference is such that their corresponding estimations of the
fractal dimension are significantly different. The apparent fractal
dimension $D_V$ obtained using the node-covering method with
equidistant integer values of $\ell$ gives $D_V=1.30$, which is the
absolute slope of the line of $\ln N_V(\ell)$ versus $\ln \ell$ for
$\ell=1, 2, 3,\cdots, 32$ (see the dotted-dashed line in
Fig.~\ref{Fig:Sierpinski:NVNE}). The regression of $\ln N_E(\ell)$
(obtained with equidistant integer values of $\ell$) as a function
of $\ln\ell$ for $\ell=1, 2, 3,\cdots, 32$, shown as the dashed
line, gives $D_E=1.37$. Both values are significantly biased
downward compared with the exact analytical value $D_f=1.5850$.
However, $D_E=1.37$ is a better estimate of $D_f$. These biases stem
from the discrete-scale invariant structure of the Sierpinski
network, which leads to geometrically increasing plateaus, as can be
seen in Fig.~\ref{Fig:Sierpinski:NVNE}. The estimated dimension
$D_E$, while also biased, is closer to the exact value, because the
Sierpinski network is exactly self-similar in terms of edges but
only asymptotically self-similar in terms of nodes. This suggests
that the edge-covering method may give better results for
finite-sized networks.

\subsection{Log-periodic oscillations and referred scaling ratio}
\label{s2:Sierpiski:LP}

\begin{figure}[htb]
\begin{center}
\includegraphics[width=10cm]{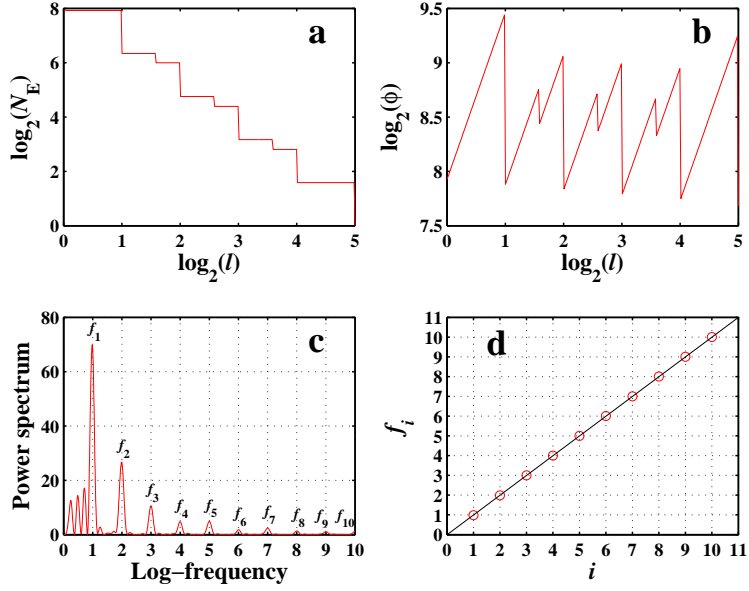}
\end{center}
\caption{Quantification of log-periodicity in the Sierpinski network.
{\textbf{a}}:
dependence of $N_E(\ell)$ defined by (\ref{Eq:NEcontinous}) as a
function of $\ell$.
{\textbf{b}}: dependence of $\phi$ as a function of $\ell$ in log-log scale,
showing strong visual evidence of the
log-periodic oscillations. {\textbf{c}}: Power
spectrum of the function $\log_2[\phi(\log_2 \ell)]$
of the variable $\log_2\ell$. The log-frequency of the main peak is $f_1=
0.985$ corresponding to the preferred scaling ratio is $\lambda= 2.02$.
{\textbf{d}}: dependence of the log-frequencies $f_i$ corresponding to
the successive peaks shown in panel {\textbf{c}} as a function of
their index $i$.
The linear dependence suggests that the succession of peaks in panel
{\textbf{c}}
corresponds to the different harmonics of $f_1$. A new estimation of $f_1$ is
obtained by the regression $f_i = i \times f_1$, which yields
$f_1= 1.002$ corresponding to a preferred scaling ratio
$\lambda= 1.997$.} \label{Fig:Sierpinski:Lomb}
\end{figure}

The data in Fig.~\ref{Fig:Sierpinski:NVNE} not only exhibits clear
scaling behavior. In addition, there are log-periodic oscillations
decorating the leading power law behavior, as expected from
the theoretical considerations presented in Section \ref{s2:ECBC}.
We now show how this log-periodicity can be extracted empirically
from the data. To extract the log-frequency $f$ and the preferred scaling ratio
$\lambda$, we first detrend $N_E$ by removing the power law to investigate
the quantity
\begin{equation}
\ln \phi(\ln(\ell)) = \ln N_E(\ell)- D_E \ln \ell~,
  \label{Eq:phi}
\end{equation}
where $D_E$ is estimated as the negative of the slope of the linear
regression of $\ln(N_E)$ as a function of
$\ln(\ell)$. It is convenient to consider the
right-side continuous function $N_E(\ell)$ for $1\leqslant \ell
\leqslant 32$ defined as
\begin{equation}
  N_E(\ell) = N_E(i),~{\mathrm{if}}~~i\leqslant\ell<i+1~,
  \label{Eq:NEcontinous}
\end{equation}
where $i=1,2,\cdots,32$. The dependence of $N_E(\ell)$ as a function
of $\ell$ is shown in Fig.~\ref{Fig:Sierpinski:Lomb}{\textbf{a}}.
The dependence of $\ln_2 \phi$ obtained from (\ref{Eq:phi})
as a function of $\ln_2 \ell$ is
shown in Fig.~\ref{Fig:Sierpinski:Lomb}{\textbf{b}}.

In order to calculate the power spectrum of the function
$\log_2[\phi(\log_2(\ell))]$ of the variable $\log_2\ell$,
we sample on 300 values with regularly spacing
in $[\log_2(1),\log_2(32)]$. The resulting power spectrum is shown in
Fig.~\ref{Fig:Sierpinski:Lomb}{\textbf{c}}. The log-frequency (in logarithm
with base $2$)
associated with the highest peak is $f_1=0.985$ and the preferred
scaling ratio is therefore $\lambda=e^{(\ln 2)/f}=2^{1/f}=2.021$.
The height of the highest peak is $69.9$, which together with the
number of data points gives a probability of false-positive for
log-periodicity essentially $0$, at all confidence levels
\cite{Zhou-Sornette-2002-IJMPC}.
In addition to the main peak in panel {\textbf{c}} of
Fig.~\ref{Fig:Sierpinski:Lomb},
many secondary peaks can be observed. They are found to be equidistant
to a very good approximation, suggesting that the corresponding log-frequencies
are nothing but the harmonics $f_2=2f_1, f_3=3f_1, \cdots, f_i=i
\times f_1, \cdots$
of the fundamental
log-frequency $f_1$. The existence of a set of harmonics is known
to increase the evidence of log-periodicity \cite{Zhou-Sornette-2002-PD}.
In addition, the measurements of the harmonic log-frequencies allows us
to improve the estimation of the
fundamental log-frequency $f_1$ by using a linear regression of $f_i$
as a function of $i$ \cite{Zhou-Sornette-2002-PD},
as shown in Fig.~\ref{Fig:Sierpinski:Lomb}{\textbf{d}}. The slope
of this regression gives
$f= 1.002$ and the corresponding preferred scaling ratio is $\lambda= 1.997$,
which are very close to the exact values $f_1=1$ and $\lambda=2$
respectively for the
Sierpinski network.

\subsection{How to use log-periodicity to improve the estimation of
the fractal dimension}
\label{s2:Method:DE}

Note that a regression of $\ln N_E(\ell)$ versus $\ln \ell$ for
equidistant integer values of $\ell=1, 2, 3, 4, 5, \cdots, 32$ gives
a biased estimate of the fractal dimension $D_E=1.37$, which is
significantly smaller than the exact value $D_f=1.5850$. In
contrast, sampling $\ell$ with a uniform logarithmic spacing as in
Sec.~\ref{s2:Sierpiski:LP} (with 300 points) gives $D_E=1.52$, much
closer to the exact value.

More specifically, we stress the general property that a fractal
system with discrete
scale invariance is better sampled using  data points at $\ell =
\ell_0 \lambda^k$, where $\ell_0$ is associated with a well-chosen
phase of the log-periodicity and $k=0,1, 2, 3 \cdots$, which are integer
powers of the underlying discrete scale factor $\lambda$. In this way,
the sub-dominant log-periodic corrections to the scaling are removed.

Using this method for sampling $N_E(\ell)$, i.e., regressing $\ln
N_E(1)$, $\ln N_E(2)$, $\ln N_E(4)$, $\ln N_E(8)$, $\ln N_E(16)$,
and $\ln N_E(32)$ for the Sierpinski network $\Delta_6$ as a
function of $\ell=1$, 2, 4, 8, 16, and 32, gives $D_E=1.5850$ which
is indeed the exact fractal dimension $D_f$.

Thus, we decrease the bias in the estimate of the scaling exponent $D_E$
when going from equidistant integer values of $\ell$ to geometrically
sampled $\ell$. Moreover, we eliminate completely the bias when the
geometrical ratio used in the sampling method is equal to the preferred
scaling ratio $\lambda$ characterizing the discrete scale invariant
structure of the network.

Obviously, this method applies and
work for other fractals only if they exhibit the
symmetry of discrete scale invariance.
In our experience, using
$\ell_0=1$ is a good choice.

\section{Application to cellular networks}
\label{s1:SF}

\subsection{The data sets}
\label{s2:CFdata}

We use the ERGO (formerly WIT) database, which provides links to
information about the functional role of enzymes. More precisely, the
ERGO database of cellular networks considers the cellular functions
divided according to bioengineering principles containing data sets for
intermediate metabolism and bioenergetics (core metabolism), information
pathways, electron transport, and transmembrane transport. We revisit
the forty three cellular networks, which have been studied in a recent
analysis developed from the point of view of scale-free networks
\cite{Jeong-Tombor-Albert-Oltavi-Barabasi-2000-Nature}.

\subsection{Log-periodic oscillations in cellular networks}
\label{s2:LPinCF}

\begin{figure}[htb]
\begin{center}
\includegraphics[width=10cm]{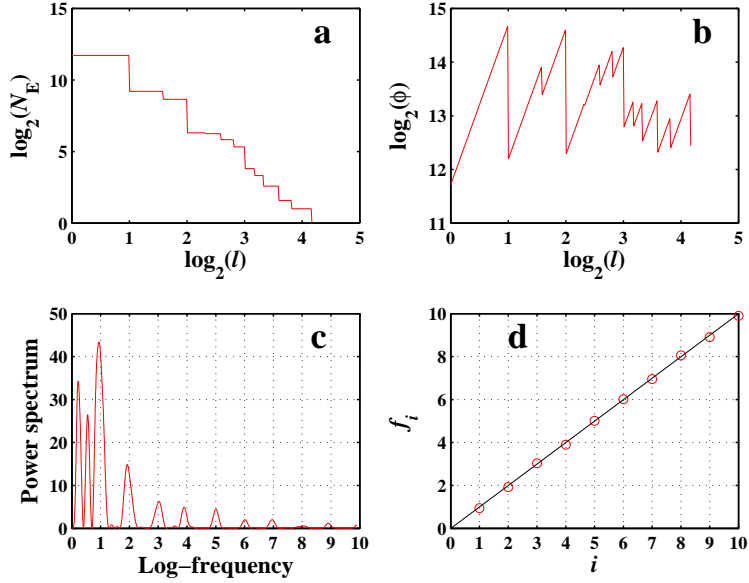}
\end{center}
\caption{Log-periodicity in the {\textit{Aquifex aeolicus}} network.
{\textbf{a}}, The dependence of $N_E(\ell)$ as a function of $\ell$.
{\textbf{b}}, The dependence of $\phi$ as a function of $\ell$. The
log-periodic oscillations are very clear. {\textbf{c}}, Power
spectrum of $\log_2[\phi(\log_2(\ell))]$. The log-frequency is $f_1=
0.9413 $ and the preferred scaling ratio is $\lambda= 2.0884 $. The
probability of false alarm is  7.2\%. {\textbf{d}}, The dependence
of $f_i$ as a function of $i$. The log-frequency is $f= 0.9990 $ and
the preferred scaling ratio is $\lambda= 2.0014 $.}
\label{Fig:CF:Lomb:AA}
\end{figure}

We consider the same 43 cellular networks analyzed in
\cite{Jeong-Tombor-Albert-Oltavi-Barabasi-2000-Nature}. We follow
the procedure described in Sec.~\ref{s1:Sierpinski} to investigate
each of these 43 cellular networks. Specifically, we implement
the edge-covering method with the
simulated annealing algorithm to find the minimal number $N_E(\ell)$
of boxes of size $\ell$ that cover all the
edges of a given network. We find that there is a power law dependence of $N_E(\ell)$
upon $\ell$ with exponent $-D_E$. In order to extract the best minimally
biased estimate of $D_E$, following the recommendations
of the previous section, we first prune
$N_E(\ell)$ by using Eq.~(\ref{Eq:NEcontinous}). We then
perform a logarithmically evenly
spaced sampling on 300 values.
Then, we detrend $N_E(\ell)$ by the power law $\ell^{-D_E}$ to
obtain $\phi(\ell)$, as shown in
Fig.~\ref{Fig:CF:Lomb:AA}{\textbf{b}}. Fourier transform is adopted
to obtain the power spectrum of $\log_2(\phi)$ with $\log_2(\ell)$
being the independent variable, as shown in
Fig.~\ref{Fig:CF:Lomb:AA}{\textbf{c}}. A clear peak at about $f=1$
as well as many remarkable harmonic peaks at $f_i=if$ are observed
for all the cellular networks. For several networks, the peak at
$f \approx 1$ is not the highest peak. However, the harmonic peaks
give strong evidence that $f \approx 1$ is the fundamental
log-frequency. In addition, the high peaks at $f \approx 0.25$
are nothing but a Nyquist frequency (in log-scale) corresponding to
the lower frequency approximately equal to half the inverse of the
whole interval (in log-scale). The total span of the data is about $\log_2 18$
and its low-frequency is $1/\log_2 18\approx 0.24$, known as the most
probable log-frequency
\cite{Huang-Johansen-Lee-Saleur-Sornette-2000-JGR}. The averaged
log-frequency over all networks is thus $f=0.97 \pm 0.05$ and the
average preferred scaling ratio if $\lambda =
{\rm{e}}^{\log(2)/f}=2.05\pm0.07$. We also plot $f_i$ versus
$i$ to achieve a better estimate of the fundamental log-frequency,
as shown in Fig.~\ref{Fig:CF:Lomb:AA}{\textbf{d}}. We see that all
plots show excellent linearity. This gives an averaged log-frequency
over all networks is thus $f=0.996\pm0.004$ and the average
preferred scaling ratio if $\lambda =
{\rm{e}}^{\log(2)/f}=2.005\pm0.005$.

We now present two additional
methods to obtain the averaged log-frequency.

\begin{figure}[htb]
\begin{center}
\includegraphics[width=8cm]{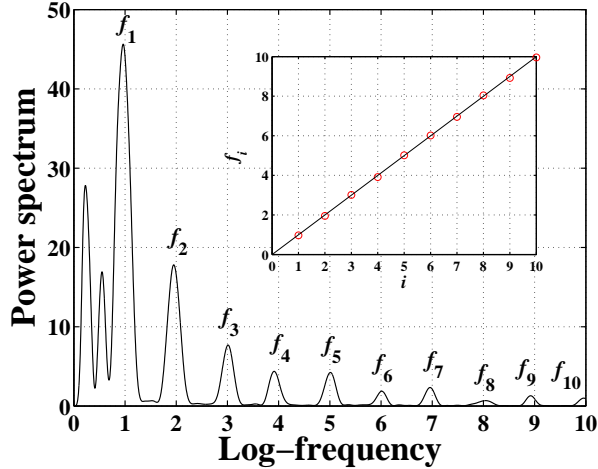}
\end{center}
\caption{Power spectrum obtained by canonical averaging over all
power spectra for 43 cellular networks. Inset: Estimation of the
fundamental log-frequency.} \label{Fig:CF:Lomb:Ave}
\end{figure}

The first method consists in a canonical analysis of the
log-periodic oscillations in which we perform averaging over all
individual power spectra to get an averaged spectrum. The idea was
initially introduced in the analysis of log-periodicity in
two-dimensional turbulence \cite{Johansen-Sornette-Hansen-2000-PD}.
Figure \ref{Fig:CF:Lomb:Ave} shows the averaged power spectrum over
43 cellular networks. The highest peak is located at $f_1 = 0.965$
(thus $\lambda = 2.051$). Again we see a number of harmonic peaks at
$f_i=if$. Linear regression of $f_i$ against $i$, shown in the inset
of Fig.~\ref{Fig:CF:Lomb:Ave}, gives a nice line with slope
$f=1.0001$ (thus $\lambda = 1.9998$), which is the fundamental
log-frequency.

The second method is to average $N_E(\ell)$ for each $\ell$ over the
43 networks, which is then analyzed as if it is from an individual
network. The highest peak is located at $f_1 = 0.989$ (thus $\lambda =
2.015$). Again we observe a number of harmonic peaks at $f_i=if$.
Linear regression of $f_i$ against $i$, shown in the inset of
Fig.~\ref{Fig:CF:Lomb:Ave}, gives a nice line with slope $f=0.999$
(thus $\lambda = 2.002$), which is the fundamental log-frequency.

\subsection{The significance of log-periodicity}

There are well established methods to assess the statistical
significance of log-periodic oscillations according to the data
points used in the analysis and the Lomb peak height
\cite{Zhou-Sornette-2002-IJMPC}.

Here, in order to test for the possibility that
the log-periodic oscillations could be artifactual, we adopt the
following bootstrapping approach. For each cellular
network, we obtain the minimal number of boxes needed to cover all
the edges of the network using simulated annealing. Due to the
self-similarity of the network, we can calculate the residuals
$\phi$ according to Eq.~(\ref{Eq:phi}). Then we reshuffle the
residuals series randomly to get a new reshuffled residuals series
$\phi'$ and then put it back to construct a new series of
edge-covering box numbers $N'_E = \phi'\ell^{-D_E}$.
Then we use the same
procedure described in Sec.~\ref{s2:LPinCF} to construct the power
spectrum. We extract the maximum height $P_m$ of the part with
log-frequency between $[0.9,1.1]$. This procedure is repeated for
1000 times, which gives 1000 values of $P_m$. For the real cellular
network, we have its counterpart height $P_N$. Finally, the
probability of a false alarm for log-periodicity
(so-called ``false positive'' or error of type II)
is defined by
\begin{equation}
  \Pr = \frac{\#[P_m>P_N]}{1000}~,
  \label{mgjmmvlwa}
\end{equation}
where $\#[P_m>P_N]$ counts the number of networks whose Lomb peak
height is larger than $P_N$.

The resulting probabilities of a false alarm for each of the 43
cellular networks are ordered by increasing values: 0.0040; 0.0060;
0.0080; 0.0090; 0.0120; 0.0140; 0.0160; 0.0160; 0.0170; 0.0170;
0.0180; 0.0210; 0.0220; 0.0220; 0.0270; 0.0300; 0.0330; 0.0390;
0.0430; 0.0510; 0.0510; 0.0580; 0.0620; 0.0700; 0.0710; 0.0720;
0.0730; 0.0770; 0.0790; 0.0830; 0.0840; 0.0880; 0.0910; 0.1010;
0.1110; 0.1120; 0.1230; 0.1500; 0.1630; 0.1640; 0.1660; 0.2250;
0.2360. We find that 19 out of the 43 cellular networks have
significant log-periodicity above the confidence level of 95\%
($\Pr<0.05$) and 33 out of the 43 cellular networks have significant
log-periodicity at the confidence level of 90\% ($\Pr<0.10$).
However, there are 10 cases that the null hypothesis that the
log-periodicity is an artifact can not be rejected even at the
significance level of 10\%.

\subsection{A simple artifactual contribution to the observed log-periodicity}

The reported probabilities of a false positive for log-periodicity
are not very small, which cast doubts on whether log-periodicity
really exists. Another curious fact is the closeness of the measured
scaling factor $\lambda$ to the number $2$. This last fact suggests
an artifact. Actually, there is a simple mechanism to produce peaks
at $\lambda \approx 2$ in finite networks (the effect disappears
eventually in the log-periodic spectral analysis for infinitely
large networks). The mechanism is based on the fact that the edges
are discrete and $\ell$ is an integer. Since $\ell$ takes integer
values, the two smallest values $\ell=1$ and $\ell=2$ allow to form
the ratio $2$. Then, the next two values $\ell=3$ and $4$ gives the
ratios $3/2$ and $4/2=2$. Combined with the two first values
$\ell=1, 2$, the ``harmonic'' $4/1$ appear. We thus see that two
approximate log-periodic oscillations appear when $N_E(\ell)$ is
sampled at intermediate real values (we use typically 300 values of
geometrically spaced $\ell$'s over a range of scale roughly $18$),
just from the existence of discreteness in the first four values of
$\ell$. This effect can be approximately observed in
Fig.~\ref{Fig:CF:Lomb:AA}{\bf{b}}: the two first oscillations are
well-formed up to $\log_2(\ell)=2$, i.e., for $\ell$ up to the value
$4$ as just described. In other words, any power law function which
is sampled on integers will exhibit some observable log-periodicity
with two to three approximate oscillations, just as a result of
discretization for the first few integer values of the scale. This
idea is confirmed by our synthetic bootstrap tests which, in
complete absence of log-periodicity (which is destroyed by the
random reshuffling), nevertheless produce peaks $P_m$ in the
log-periodic spectral analysis which are often comparable to those
observed in the 43 networks, hence the relatively large values of
$\Pr$ (defined in Eq.~(\ref{mgjmmvlwa})) obtained in the previous
section.

We conclude this rather pessimistic argument (with respect to
the existence of genuine log-periodicity) by two positive notes.
\begin{itemize}
\item First, notwithstanding this artificial effect, one cannot deny
that the real cellular networks
have often higher Lomb peaks for log-periodicity than the reshuffled ones,
suggesting other sources of discrete scale invariance, perhaps genuine.
If this is the case, the
value $\lambda \approx 2$ remains to be explained. The simplest
argument could be that the
dichotomous discrete hierarchy is perhaps the most natural and most robust
discrete hierarchy that can be found.

\item Second, as we have shown in section 2, taking into account
the presence of log-periodicity, spurious or not, to sample $N_E(\ell)$
offers a priori a better less-biased estimator for the fractal
dimension $D_E$.
\end{itemize}

\subsection{Fractal dimensions of cellular networks}

\begin{figure}[htb]
  \begin{center}
  \includegraphics[width=8cm]{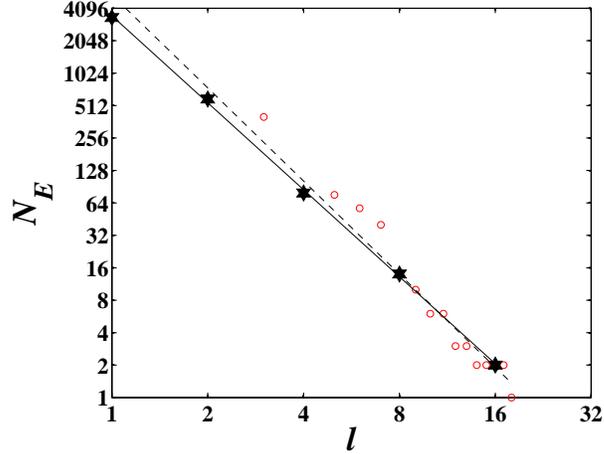}
  \end{center}
\caption{Determination of the fractal dimensions. The solid line is
a fit to the data points at $\ell = 2^k$ using Method 1. The dashed
line is the fit to all data points with Method 3. The open circles
are all the calculated values of $N_E$ for all integer $\ell$'s while the stars
are sampled geometrically at $\ell_k = 2^k$.
}
\label{FigCF:DE:AA}
\end{figure}

We revisit in this section the estimation of fractal dimensions of
the 43 cellular networks using four different methods described
below. Note that all these methods are based on the edge-covering
method.
\begin{itemize}
   \item {\textbf{Method 1.}} We have shown in Sec. \ref{s2:LPinCF} that
there are significant log-periodic oscillations in all the 43
cellular networks with a universal preferred scaling ratio $\lambda
= 2$. We can the use the data points at $\ell=2^k$: $N_E(1)$,
$N_E(2)$, $N_E(4)$, $\cdots$, shown in Fig.~\ref{FigCF:DE:AA} as
stars. The inverse slope of $\ln[N_E(2^k)]$ against $\ln (\ell_k)=k \ln 2$ is an
estimate of the fractal dimension. This method is illustrated in
Fig.~\ref{FigCF:DE:AA}.
   \item {\textbf{Method
2.}} This method uses a logarithmic sampling using
Eq.~(\ref{Eq:NEcontinous}). For each network, 300 evenly spaced data
points in the logarithmic abscissa are used and its inverse slope of
the data in log-log plot is an estimate of the fractal dimension.
   \item {\textbf{Method 3.}} This method uses all the data point at
$\ell=1,2,3,\cdots$, shown as circles in Fig.~\ref{FigCF:DE:AA}, and
fit it in log-log plot. The inverse of the slope is taken as the
estimate of the fractal dimension. This method is also illustrated
in Fig.~\ref{FigCF:DE:AA}.
   \item {\textbf{Method 4.}} This method is the same as {\textbf{Method 3}}
except that this method plot $\ln[N_E(\ell)]$ versus $\ln(\ell+1)$, as
used in \cite{Song-Havlin-Makse-2005-Nature}.
\end{itemize}

\begin{figure}[htb]
\begin{center}
\includegraphics[width=8cm]{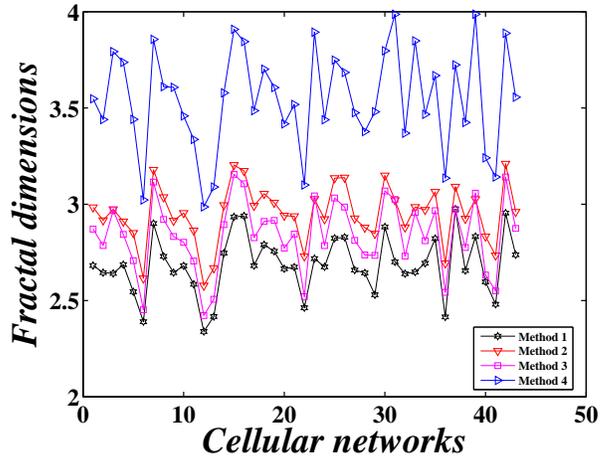}
\end{center}
\caption{Comparison of the fractal dimensions of the 43 cellular
networks estimated using four different methods.} \label{Fig:CF:DE}
\end{figure}

We calculated the fractal dimensions of the 43 cellular networks
using these four different methods. The results are shown in
Fig.~\ref{Fig:CF:DE}. The averages of the estimated fractal
dimensions are $D_E=2.68 \pm 0.15$ (method 1), $D_E=2.95\pm 0.15$
(method 2), $D_E=2.84
\pm 0.19$ (method 3), and $D_E=3.47 \pm 0.27$ (method 4), respectively. It is interesting
to note that $D_E=3.54 \pm 0.27$ for Method 4 is identical to
$D_V=3.5$ estimated using node-covering
\cite{Song-Havlin-Makse-2005-Nature}. We argue that our Method 1
based on the log-periodic sampling gives a better and more
robust estimation of the fractal
dimension, due to a minimal bias. We believe that our method
of edge-covering with simulated annealing and log-periodic sampling
minimizes the significant bias in the log-log regression.

\section{Conclusion}

In summary, we have proposed a new box-counting method for networks,
in which boxes are tiled to cover all the edges of the network. The
minimal number $N_E$ of boxes of size $\ell$ in the edge-covering
method is obtained by a simulated annealing algorithm. We have
performed detailed analytical and numerical analysis of the
Sierpinski network. The results show that the Sierpinski network is
strictly self-similar only when using the edges (and not the nodes).
We have shown that the node-covering method gives a stronger
downward bias estimate of the fractal dimension of the Sierpinski
network.

In addition, we have shown that we can use
the discrete scale invariance of the Sierpinski network to
characterize the log-periodic
oscillations in $N_E(\ell)$ and develop a method which removes
completely the bias in the estimation of the fractal dimension $D_E$.
Taking into account the presence of log-periodicity
to adapt the sampling of the function $N_E(\ell)$
allows us to design a better estimator for the fractal
dimension of self-similar networks.

We have applied this improved method to 43 cellular networks
previously studied in the literature. We found that $N_E$ scales
with respect to $\ell$ as a power law $N_E \sim \ell^{-D_E}$ with
the fractal dimension $D_E=2.67\pm0.15$. Some of the cellular
networks exhibit log-periodic oscillations in $N_E$. However, a
bootstrapping statistical test shows that the existence of
log-periodicity in cellular networks is not fully conclusive.

\bigskip
{\textbf{Acknowledgments:}} This work was supported by the National
Natural Science Foundation of China (Grant No. 70501011) and the Fok
Ying Tong Education Foundation (Grant No. 101086).

\bibliography{E:/papers/Bibliography}

\end{document}